\documentclass{article}
\usepackage{dcase2023,amsmath,graphicx,url,times,booktabs, tabularx}
\usepackage{xcolor} 
\usepackage{cite}
\usepackage{color}
\usepackage[font=footnotesize,labelfont=bf,footnotesize]{caption}

\title{Two vs. Four-Channel Sound Event Localization and Detection}

\name{Julia Wilkins$^1$, Magdalena Fuentes$^1$, Luca Bondi$^2$,
      }
\secondlinename{	  
      Shabnam Ghaffarzadegan$^2$, Ali Abavisani$^2$, Juan Pablo Bello $^1$
      }
\address{$^1$ New York University, New York, NY, USA, \\          
        $^2$ Bosch Research, Pittsburgh, PA, USA \\
        \ninept{\url{jw3596@nyu.edu}} \\
        \ninept \textcolor{magenta}{\url{https://github.com/juliawilkins/SELD-2v4-DCASE23/}}
        \vspace{-0.5cm}
}

\begin{document}

\ninept
\maketitle

\begin{sloppy}

\begin{abstract}
{Sound event localization and detection (SELD) systems estimate both the direction-of-arrival (DOA) and class of sound sources over time. In the DCASE 2022 SELD Challenge (Task 3), models are designed to operate in a 4-channel setting. While beneficial to further the development of SELD systems using a multichannel recording setup such as first-order Ambisonics (FOA), most consumer electronics devices rarely are able to record using more than two channels. For this reason, in this work we investigate the performance of the DCASE 2022 SELD baseline model using three audio input representations: FOA, binaural, and stereo. We perform a novel comparative analysis illustrating the effect of these audio input representations on SELD performance. Crucially, we show that binaural and stereo (i.e. 2-channel) audio-based SELD models are still able to localize and detect sound sources \textit{laterally} quite well, despite overall performance degrading as less audio information is provided. Further, we segment our analysis by scenes containing varying degrees of sound source polyphony to better understand the effect of audio input representation on localization and detection performance as scene conditions become increasingly complex. 
}
\end{abstract}


\section{Introduction}

Sound Event Localization and Detection (SELD) is the process of estimating the direction-of-arrival (DOA) and class of sound events over time, given an input audio signal. SELD systems can translate well to a variety of real-world applications, including navigation for autonomous systems and assistive robotic devices. 
SELD methods are rooted in traditional signal processing techniques for multichannel audio processing, such as Steered Response Power~\cite{dibiase2000high} and acoustic intensity vectors~\cite{perotin2018crnnIV}. For human-inspired audio recordings (e.g. binaural recordings), interaural time difference (ITD) and interaural level difference (ILD)  are commonly used  to characterize the direction of arrival of sounds \cite{stern2006binaural}. However, these cues alone have shown limitations in terms of localization accuracy in real-world scenes that are particularly noisy, reverberant, or polyphonic \cite{giguere1993sound, hafezi2017augmented, evers2014multiple}. Deep learning approaches were recently popularized to address these challenges in the context of SELD tasks \cite{2018baselinemodel,gulati2020conformer,Zhao2021,Gan_2019_ICCV, Grumiaux2021ASO}; most systems still utilize signal processing-based features like generalized cross correlation (GCC) and Mel spectrograms but benefit from automatic feature learning to improve robustness in difficult scene conditions \cite{knapp1976generalized, Grumiaux2021ASO,he2018deep, 2018baselinemodel}. For example, in \cite{adavanne2018direction}, authors use a CRNN architecture with magnitude and phase spectrograms from multichannel audio to show accurate DOA estimation and multiple sound source detection in reverberant conditions.

In the DCASE 2022 SELD challenge (Task 3), models were evaluated using real multichannel sound recordings. Participants had access to real recordings for development and could also use additional synthetic or real data for training.
The challenge operates in a multichannel setting, utilizing two formats of 4-channel recordings: first-order Ambisonics (FOA) and a tetrahedral mic array. We are interested in exploring the capabilities of current SELD systems using more commonly found 2-channel microphone setups, namely binaural and stereo, as typical consumer electronics devices lack such complex 4-channel configurations.

There is little prior research quantifying the effect of using various audio input representations (i.e. 2 vs. 4-channel audio) for SELD tasks in deep learning-based systems. In the psychoacoustics community, this effect is well-studied; it is known that there is a general loss in spatial understanding between 4-channel audio configurations (e.g. Ambisonics) vs. 2-channel configurations (e.g. binaural or stereo). \cite{wightman1989headphone, blauert1997spatial}. Humans can localize lateral sound sources well in binaural and stereo settings, but front-back confusion may increase without sufficient spatial information \cite{stern2006binaural, rudzki2019auditory, thresh2017direct}. Further, perceiving the elevation of sound sources when listening to stereo audio in particular has been shown to be very difficult, largely due to the lack of interaural cues present in this recording configuration unlike that of a binaural setup \cite{blauert1997spatial}. However, these phenomena are underexplored in the context of deep learning-based systems for SELD. In \cite{adavanne2018multichannel}, authors compared sound event detection performance using synthetic FOA, binaural, and monaural audio data in a CRNN-based system. Our approach differs significantly in that we provide a quantitative analysis of localization \textit{and} detection performance, we use a FOA dataset of real recordings in addition to synthetic and decode these recordings to binaural, and lastly we include the stereo audio configuration as a point of comparison as this is common in consumer electronics devices today.

In this work we present a novel comparative analysis of the DCASE 2022 SELD baseline model across FOA, binaural, and stereo audio input representations. To the best of our knowledge, this is the first work quantifying the effect of these audio configurations on both localization and detection performance in a deep-learning based SELD system. We show that lateral sound source localization remains fairly accurate in the 2-channel settings despite an overall degradation in SELD performance, and provide an analysis of performance in scenes of varying levels of polyphonic sound source complexity.

\vspace{-1em}
\section{Problem Formulation}
In this manuscript, we examine the problem of Sound Event Localization and Detection (SELD) under different audio input representations: first-order Ambisonics (FOA), binaural, and stereo recordings.
In this context, \textit{detection} refers to determining the number of active sound sources per class over time, while \textit{localization} aims at identifying the azimuth and elevation angle for each of the active sources over time.
While Ambisonics recordings provide state-of-the-art performance in SELD \cite{9616002}, in practical applications we hypothesize that binaural and stereo recordings are more accessible.

We rely on the most popular framework used by participants in the DCASE 2022 Challenge Task 3. A multichannel audio recording is fed as input to a Convolutional Neural Network (CNN), whose output is a 4-dimensional matrix arranged according to the Multi-Activity Coupled Cartesian DOA (ACCDOA) format~\cite{multiaccdoaShimada2022}. For a given class, time instant, and sound source index, the model arrives at a three-dimensional vector $(x,y,z)$ whose orientation represents the direction of arrival of the sound, and whose intensity is directly proportional to the likelihood of a sound of that class being present at a given time.

\textbf{First-order Ambisonics (FOA)}:
FOA is a 4-channel, 3D audio recording format. In FOA, each channel corresponds to a spherical harmonic component representing a change in sound pressure in a specific direction \cite{gerzon1973periphony}. The channels \textit{W, Y, Z, X} map to the omnidirectional, left-right, vertical, and front-back directions of sound pressure change, respectively.

\textbf{Binaural}:
The binaural recording technique aims to capture 3D audio in just two channels, ideally simulating the experience of a human experimncing auditory cues. Binaural audio is typically recorded using two microphones placed in the ears of a dummy head (e.g. Neumann KU100), or synthesized using the head-related transfer functions (HRTFs) of such a dummy head \cite{engel2022assessing}.
Binaural recordings deliver immersive spatial sounds containing amplitude, time and timbral differences of two channels vs. traditional stereo recordings where only amplitude and time differences are available.

\textbf{Stereo}:
In stereo recordings, two microphones are used to capture the left and right audio channels independently. This differs from binaural recordings; in the binaural configuration the goal is to simulate a human's listening experience. Critically,  in a stereo setup, elevation differentiation cannot be perceived; binaural recordings contain the filtering effect of the head, ear pinna, and torso and this is not present in a stereo recording configuration \cite{blauert1997spatial}.

\section{Experimental Setup}

\begin{figure*}[h]
\centering
\includegraphics[width=0.82\textwidth, trim={0 0cm 0cm 0cm}]{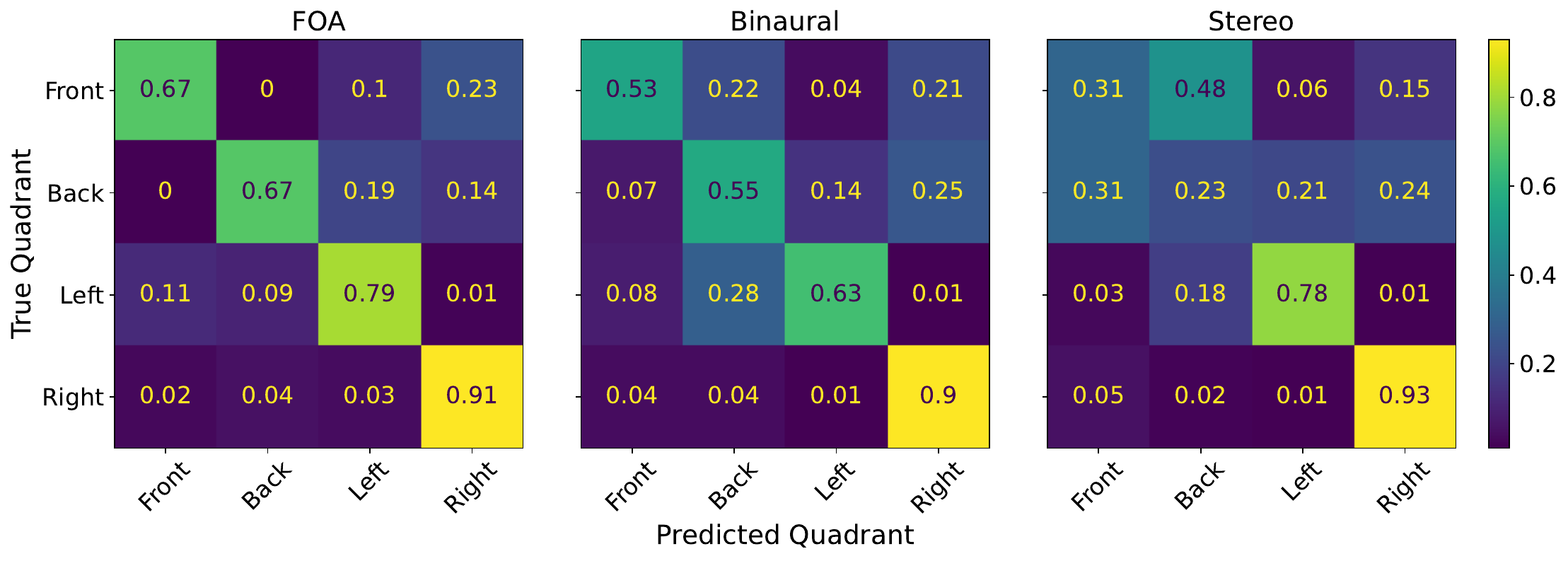}
\caption{\footnotesize{Normalized confusion matrices showing true vs. predicted quadrant of sources across audio configurations. The FOA model performs near-perfect at distinguishing front and back sources, while front and back sources are commonly confused in binaural and stereo settings. Quadrants of size $90^{\circ}$ are defined based on the azimuth angle of a sound source:  Front $\in[-45^{\circ},45^{\circ}]$, Left $\in[45^{\circ},135^{\circ}]$, Back $\in[135^{\circ},\pm180^{\circ}] \cup [\pm180^{\circ},-135^{\circ}] $, Right $\in[-135^{\circ},-45^{\circ}]$ }}
\label{fig:confmatrix}
\vspace{-1em}
\end{figure*}

\subsection{Datasets}
Following the setup of the DCASE 2022 Task 3 challenge, we rely on the STARSS22 dataset~\cite{starrs_dataset}, together with a synthetic mixture (SYNMIX) for baseline training\footnote{https://zenodo.org/record/6406873\#.Y\_-SBuzMK2o.} provided by the organizers of the challenge. The STARSS22 dataset is comprised of 121 recordings of various lengths of real sound scenes across 13 sound event classes, with around 5 hours of audio recordings in 4-channel FOA format and an interpolated tetrahedral microphone array. At the time of this work, the evaluation set was not yet released, so we use the ``development" partition of train and test, consisting of 67 and 54 recordings, respectively. The dataset contains instances with up to 5 simultaneous sound sources, and up to 4 simultaneous sources of the same class, though 2-source polyphony is much more frequent.

Due to the small size of the STARSS22 dataset, a base set of synthetic data was also provided to participants (SYNMIX). This data is synthesized using audio samples from FSD50k \cite{fonseca2021fsd50k} convolved with Spatial Room Impulse Responses from the TAU-Nigens Spatial Sound Events 2020 \cite{2020taunigens} and 2021 \cite{2021taunigens}. The dataset contains 1200, 1-minute synthesized FOA recordings across classes mapped to the classes present in STARSS22, and maximum polyphony of 2 sources.

Both datasets are annotated at 100ms resolution with labels of sound source class, azimuth, and elevation as well as additional flags for overlapping sound events. The azimuth angles $\phi \in [-180^{\circ} , 180^{\circ} ]$, and elevation $\theta \in [-90^{\circ} , 90^{\circ} ]$, with $0^{\circ}$ at front. Note that azimuth angles increase counterclockwise.

\vspace{-0.5em}
\subsection{Input representations}
To fairly compare the three multichannel audio representations, we look at the problem of sound localization on the horizontal plane only by removing the elevation component, thus fixing elevation to $0^{\circ}$ in the ground truth.
We train and test separately for each input representation using the same acoustic scenes, simply replacing the original FOA audio representation with binaural or stereo audio, as per following procedures.

\textbf{FOA $\rightarrow$ Binaural}:
To decode the original FOA audio from the STARRS22 and synthetic datasets to binaural, we used the BinauralDecoder plug-in from the IEM Plug-In Suite\footnote{https://plugins.iem.at/docs/plugindescriptions/\#binauraldecoder.}.
This decoder uses pre-processed Neumann KU100 dummy head HRTFs via the magnitude least-squares (MagLS) method proposed in \cite{schorkhuber2018binaural}. 
We apply this binaural decoding to all FOA audio used in training and testing, yielding 2-channel binaural audio for our experiments \footnote{https://github.com/juliawilkins/ambisonics2binaural\_simple.}.

\textbf{FOA $\rightarrow$ Stereo}:
To convert our FOA audio to stereo, we used a very simple translation:  $ \textit{left} = W + Y $ and $ \textit{right} = W - Y $, following
~\cite{Zotter2019}. Note that $W$ is the omnidirectional signal and $Y$ is the first-order horizontal (left-right) component. An increase in air pressure from left causes an increase in values of $Y$ and an increase in pressure from the right causes a decrease in values of $Y$. Because of this, the simple translation above allows us to move easily from FOA to left and right channels yielding 2-channel stereo audio.

\vspace{-0.5em}
\subsection{Baseline model}
The model used for our analysis is the DCASE 2022 Task 3 Baseline model\footnote{https://github.com/sharathadavanne/seld-dcase2022.}. The architecture is similar to the CRNN-based model initially proposed in \cite{2018baselinemodel}, with extensions to accommodate simultaneous sources of the same class in the Multi-ACCDOA format~\cite{multiaccdoaShimada2022}. The input to the model is the multichannel audio, segmented into 5-second chunks, yielding a sequence of 50 x 0.1 second frames. In the FOA configuration, Mel spectrogram features are used to capture frequency information and intensity vectors provide spatial information. In the binaural and stereo settings, we modify the model slightly to use Mel spectrograms and GCC features. GCC features are commonly used in 2-channel localization settings to capture Time Difference of Arrival (TDOA) information between two microphones. Audio is resampled to 24kHz, and 64 Mel coefficients are computed from an STFT on windows of 1024 samples with a hop size of 480 samples. The model has 604.5K trainable parameters.
Models are trained for a multi-output regression task, with a mean-squared-error loss, for 200 epochs using 1 RTX 8000 GPU, in batches of 64 samples with a learning rate of $10^{-3}$. The model checkpoint with the lowest validation loss is selected.

\vspace{-0.5em}
\subsection{Data augmentation via Audio Channel Swapping (ACS)} \label{acs}

An initial exploration of the STARSS and SYNMIX datasets revealed that the distribution of azimuth angles across sound sources was largely imbalanced, with far more sound sources in the front and right regions than in the left and back. Following \cite{Du_NERCSLIP_task3_report}, we hypothesize that localization performance on the real test dataset could be improved by balancing this distribution. To do so, we use a data augmentation technique known as Audio Channel Swapping (ACS) \cite{augacs}. We perform 3 transformations involving azimuth to simulate the rotation of sound sources by $90^{\circ}$, $180^{\circ}$, and $270^{\circ}$. We performed different permutations of swapping and negating the $X$ and $Y$ of FOA channels directly. This simple augmentation strategy not only quadruples our overall dataset size but more importantly gives us a uniform distribution of azimuth angles. We show that this augmentation has a significant impact on localization performance in Table \ref{tab:foa_aug_results}. Please refer to \cite{augacs} for more details on ACS.

\vspace{-0.5em}
\subsection{Evaluation metrics}
We use the joint localization and detection metrics as defined by the DCASE 2022 Task 3 SELD Challenge in our proceeding analysis. The F-Score and error rate (ER) capture location-dependent detection. True Positives (TP) and False Positives (FP) are considered with a tolerance $20^{\circ}$ in the direction of arrival. Class-dependent localization error (LE) and localization recall (LR) measure localization performance without considering the spatial threshold. See \cite{politis2020overview} for more details on SELD metrics. \vspace{-0.5em}

\section{Results}\label{sec:Results}

\subsection{A baseline model for FOA input} Prior to evaluating the impact of different input representations, we first assess the performance of a baseline model trained and evaluated on FOA input using varied training data configurations. The STARSS22 and SYNMIX dataset are both quite imbalanced in terms of distribution of sound source across azimuth angles. As described in Section \ref{acs}, we use Audio Channel Swapping (ACS) to mitigate this problem and balance the distribution at train time.

Table \ref{tab:foa_aug_results} reports results for 5 training data configurations: 
\textbf{A}: training and evaluating only in azimuth using STARSS22 dataset; \textbf{B}: adding SYNMIX dataset to A's training; \textbf{C}: adding ACS augmentation to B's training, \textbf{$\text{B}^{+E}$}: training and evaluating B in both azimuth and elevation; \textbf{$\text{C}^{+E}$}: training and evaluating C in both azimuth and elevation. 
Note that $\text{B}^{+E}$ and $\text{C}^{+E}$ help us to understand the impact of removing elevation in the overall metrics. By comparing $\text{C}^{+E}$ and $\text{C}$, we see how removing elevation improves all metrics, as one could imagine given less degree of freedom in the predictions.
Moreover, we see an improvement in the joint localization and detection metrics across the board with the addition of the augmented data. Hence, we use $\textbf{C}$ as our reference configuration to assess the impact of the input representation in proceeding sections.

\begin{table}[tbh]
\centering

\begin{tabular}{ccccccccccccc}
\toprule

 \textbf{Conf.} & \textbf{SELD $\downarrow$} & \textbf{ER$\downarrow$} & \textbf{F $\uparrow$} & \textbf{LE $\downarrow$} & \textbf{LR $\uparrow$}   \\

\midrule

$\text{A}$ & 0.65 & 0.73 & 15.3\% & 53.7$^{\circ}$ & 27\%  \\
$\text{B}$ & 0.47 & 0.62 & 34.5\% & 22.5$^{\circ}$ & 51\%  \\
$\text{C}$ & 0.42 & 0.56 & 43.3\% & 16.9$^{\circ}$ & 54.1\%  \\
\midrule
$\text{B}^{+E}$ & 0.53 & 0.70 & 27.3\% & 26.1$^{\circ}$ & 47.5\% \\
$\text{C}^{+E}$ & 0.48 & 0.62 & 33\% & 22.7$^{\circ}$ & 51\% \\

\bottomrule
\end{tabular}

\caption{\footnotesize{Results with \textbf{FOA} input across different configurations; $\text{A}$: STARSS22; $\text{B}$: \text{A} + SYNMIX; $\text{C}$: \text{B} with ACS; $\text{B}^{+E}$ and $\text{C}^{+E}$: \text{B} and \text{C} are trained and evaluated using both azimuth and elevation.
Results are reported on the STARSS22 DCASE dev-test set.
$\downarrow$ indicates metrics that are better when value is lower, $\uparrow$ viceversa.
}}

\label{tab:foa_aug_results}
\vspace{-2em}
\end{table}

\subsection{Comparing audio input representations}
Table \ref{tab:results_audioconfig} reports results when changing input representation, moving from the highly-privileged FOA representation, to binaural, and stereo audio. 
Our experiments show that as one moves from FOA to binaural and stereo, overall SELD model performance degrades. While this is to be expected because binaural and stereo audio are not designed to capture full spatial audio, this is the first quantification of deep learning-based SELD performance across these audio input representations on real multichannel recordings lays the groundwork for our deeper proceeding analysis.

\begin{table}[tbh]
\centering
\begin{tabular}{ccccccccccccc}
\toprule

\textbf{Input} & \textbf{SELD $\downarrow$} & \textbf{ER $\downarrow$} & \textbf{F $\uparrow$} & \textbf{LE $\downarrow$} & \textbf{LR$\uparrow$}   \\
\midrule

FOA  & 0.42 & 0.56 & 43.3\% & 16.9$^{\circ}$ & 54.1\%  \\
 Binaural & 0.50 & 0.67 & 33.9\% & 30.1$^{\circ}$ & 49.2\%  \\
 Stereo & 0.60 & 0.76 & 21.7\% & 42.9$^{\circ}$ & 38.8\%  \\

\bottomrule
\end{tabular}
\caption{\footnotesize{Results for models trained using STARSS22 + SYNMIX using ACS, with different audio input representations. Results are reported on the STARSS22 DCASE development-test set. $\downarrow$ indicates metrics that are better when value is lower, $\uparrow$ viceversa.}}

\label{tab:results_audioconfig}
\vspace{-1em}
\end{table}

\vspace{-0.5em}
\subsection{Localization error by sound source quadrant} 
We are also interested in dissecting localization performance to understand where key success and failure points occur in terms of sound source position and polyphonic scene conditions.

In Figure \ref{fig:confmatrix}, we show a set of confusion matrices illustrating the distribution of true quadrants of sound sources vs. predicted quadrants across audio input representations. We segment the $90^{\circ}$ quadrants as follows, based on azimuth angle: Front $\in[-45^{\circ},45^{\circ}]$, Left $\in[45^{\circ},135^{\circ}]$, Back $\in[135^{\circ},\pm180^{\circ}] \cup [\pm180^{\circ},-135^{\circ}] $, Right $\in[-135^{\circ},-45^{\circ}]$. Notably, using the FOA representation, the model has near-perfect performance in terms of distinguishing front vs. back sources. In the binaural setting, we see an increase in front-back confusion, and in the stereo setting this error is glaring as 48\% of sources in the front are predicted in the back quadrant. In fact, this is a well-studied topic in psychoacoustics related to the cone of confusion phenomenon, which occurs when a sound source is equidistant to both the left and right ears \cite{rayleigh1907xii, balan2018systematic, shinn2000tori}. Thus, it is difficult for the listener to distinguish whether a sound source is in front or behind them. It is likely that our binaural model is affected by this as well. Across audio input representations, the accuracy of source detection in the left and right quadrants is fairly consistent, showing reliability in terms of lateral sound source detection given 2- or 4-channel audio input.

\begin{figure}[htb]
\centering
\includegraphics[width=0.5\textwidth, trim={0 0.5cm 0 0.5cm}]{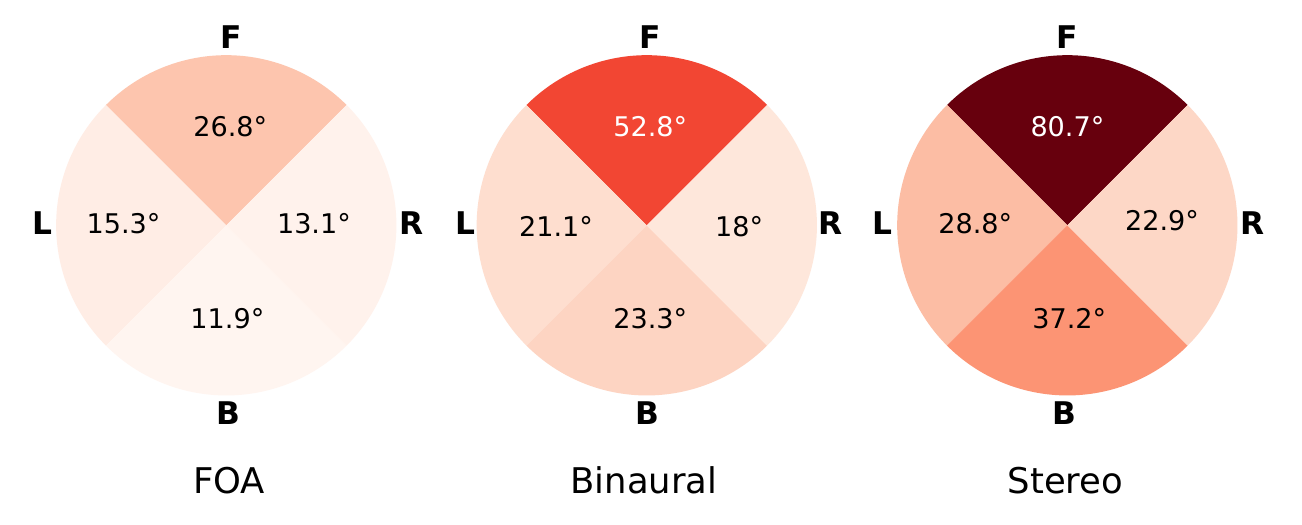}
\caption{\footnotesize{Average localization error across audio representations, based on ground truth sound source quadrant position. Results are normalized by number of instances of sound sources per quadrant.}}
\label{fig:quad}
\end{figure}
In Figure \ref{fig:quad}, we analyze average localization error (LE) based on the quadrant of the ground truth sound sources. In the FOA setting, the difference of LE between the left, right, and back quadrants is quite small, however the error for sources in the front is nearly double that of the other quadrants. In the binaural setting, LE increases in the front and back quadrants, approximately doubling that of the FOA setting, though this increase is much less notable in the lateral (left-right) regions. 
Further, in the stereo context, we find similar trends but with overall poorer performance. The front and back LE are over three times that of the FOA model, with less significant degradation in the performance of the left and right quadrants. Here, we crucially observe that despite the binaural and stereo models struggling to localize sources in the front quadrant in particular compared to the FOA system, these 2-channel models are still able to localize sources laterally quite well.

\subsection{SELD performance in polyphonic conditions}
The DCASE SELD challenge is unique in that the test dataset contains real audio recordings with multiple overlapping sound sources. Hence, investigating SELD model performance in complex polyphonic conditions can help us better understand how these systems handle more complex scene conditions that are closer to reality. In Figure \ref{fig:recall}, we analyze localization recall (LR) of the FOA, binaural, and stereo models in the presence of 1, 2, 3, and 4 simultaneous sources (this encapsulates both simultaneous sources of the same or different classes). Note that approximately $56\%$ of frames contain 1 source, $31\%$ contain 2, $10\%$ contain 3, and $3\%$ contain 4 or more simultaneous sources, so we normalize by source count accordingly in Figure \ref{fig:recall}. We show that LR steadily decreases in all audio configurations as the number of polyphonic sound sources increases in Figure \ref{fig:recall}. The model struggles to detect the correct number of sources as the scene conditions become increasingly complex, though proportionally the decrease in recall is relatively similar across audio contexts as polyphony increases. 
\begin{figure}[htb]
\centering
\includegraphics[width=0.5\textwidth, trim={0 1cm 0 0}]{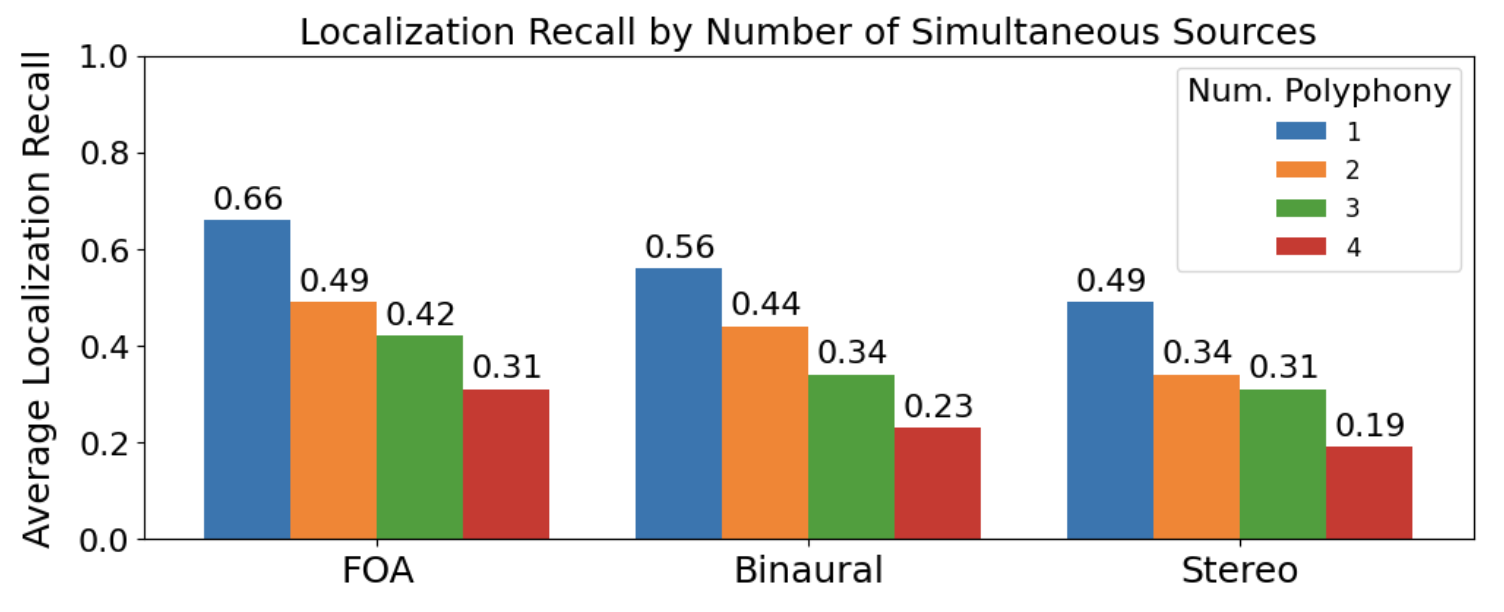}
\caption{\footnotesize{Localization recall in multiple audio representations, segmented by number of simultaneous sources in the test data and normalized by number of sources satisfying each condition.}}
\label{fig:recall}
\end{figure}

We also analyze localization error (LE) across polyphonic conditions. Here we find that while on average LE increases as we use less-informative audio representations (i.e. stereo), it is not a fully monotonically increasing trend across polyphonic conditions. In the FOA setting, the LE is similar regardless of level of polyphony. In the binaural and stereo settings, there is a much larger spread of LE across conditions, however not in a monotonically increasing manner, e.g. in the stereo setting the average LE is $31.3^{\circ}$ in the occurrence of 3 overlapping sources vs. $46.1^{\circ}$ for 2 sources. We hypothesize that there are many interacting effects contributing to this, including but not limited to class imbalance in different polyphonic conditions, simultaneous sources of the same class, and the nature of the LE metric as it does not take false negatives into account.

\section{Conclusion}
This work presents a novel comparative analysis of the DCASE 2022 SELD baseline model across first-order Ambisonics, binaural, and stereo audio input representations. We show quantitatively that while localization and detection performance decreases given less informative audio representations, binaural and stereo-based SELD models are still able to localize lateral sound sources relatively well. These findings could be highly informative in the development of applications such as an audio-visual navigation system equipped with a stereo microphone configuration and a camera; if we are confident in lateral source localization based on auditory cues, we can lean more on visual cues for sources directly in front of the camera. Future work in this space could entail an investigation into the effect of sound source class or of overlapping sources of the same class on localization performance across polyphonic conditions and audio input representations.

\newcommand\blfootnote[1]{%
  \begingroup
  \renewcommand\thefootnote{}\footnote{#1}%
  \addtocounter{footnote}{-1}%
  \endgroup
}

\blfootnote{This work is supported by NSF Grant No. IIS-1955357.}

\footnotesize
\bibliographystyle{IEEEtran}
\bibliography{refs}

\end{sloppy}
\end{document}